\documentclass{article}

\usepackage{PRIMEarxiv}

\usepackage[utf8]{inputenc} % allow utf-8 input
\usepackage[T1]{fontenc}    % use 8-bit T1 fonts
\usepackage{hyperref}       % hyperlinks
\usepackage{url}            % simple URL typesetting
\usepackage{booktabs}       % professional-quality tables
\usepackage{amsfonts}       % blackboard math symbols
\usepackage{nicefrac}       % compact symbols for 1/2, etc.
\usepackage{microtype}      % microtypography
\usepackage{lipsum}
\usepackage{fancyhdr}       % header
\usepackage{graphicx}       % graphics

\usepackage{color}
\usepackage{amsmath}
\usepackage{amssymb}
\usepackage{setspace}

\usepackage{lineno}

\usepackage{bm}
\usepackage{multirow}

\graphicspath{{media/}}     % organize your images and other figures under media/ folder

\title{A Unified Multiscale Auxiliary PINN Framework for Generalized Phonon Transport
}

\author{
  Roberto Riganti \\
  Physics Department, \\ Boston University \\
  Boston, MA, USA\\
  \And
  Luca Dal Negro\textsuperscript{*} \\
  Department of Electrical \& Computer Engineering, \\
  Division of Materials Science \& Engineering,\\
  Physics Department, and Photonics Center \\
  Boston University, Boston, MA, USA\\
  \textsuperscript{*}\texttt{dalnegro@bu.edu} \\
}

\begin{document}
\maketitle

\begin{abstract}
Nanoscale thermal transport is governed by the phonon Boltzmann transport equation (BTE). However, simulating the sub-continuum dynamics remains computationally prohibitive due to the high dimensionality of the phase space and the intrinsic nonlinearity of the scattering collision operator. Traditional numerical solvers and standard physics-informed neural networks (PINNs) inherently struggle with these integro-differential equations due to deterministic quadrature limitations, artificial thermalization introduced by the relaxation time approximation (RTA), and multiscale spectral bias. This work introduces a multiscale auxiliary physics-informed neural network (MTNet) to solve the generalized equation of phonon radiative transfer (GEPRT). By leveraging an auxiliary formulation, this mesh-free framework recasts the GEPRT into a fully differential system, enabling the analytical evaluation of scattering operators via automatic differentiation and facilitating scalable multi-GPU parallelization. To circumvent optimization stiffness, the architecture employs a decoupled, shallow neural network explicitly constrained by radiative equilibrium. MTNet is validated by simulating steady-state cross-plane transport in a silicon thin film, successfully capturing ballistic-diffusive regimes and characteristic boundary slips across extreme temperature gradients ($\Delta T = 100$ K) beyond the standard linearization approach. Furthermore, we show that our framework successfully solves a geometric inverse problem in a slab geometry, retrieving the unknown slab thickness based only on interface temperature constraints in the mesoscopic regime. Ultimately, MTNet establishes a robust, fully differentiable foundation for predicting high-fidelity kinetic transport and extracting material properties in next-generation nanostructures.
\end{abstract}

% keywords can be removed
\keywords{Neural Networks \and Thermal Transport \and Multiscale \and Inverse Problems}

\section{\label{sec:intro}Introduction}
As the miniaturization of electronic and optoelectronic devices approaches the nanometer regime, traditional macroscopic models of heat conduction, such as Fourier's law, increasingly fail to accurately predict thermal behavior~\cite{zhang_nanomicroscale_2020, chen_nanoscale_2023}. At the microscopic length scale, the physical dimensions of the system become comparable to or smaller than the mean free path of the primary heat carriers, such as phonons in semiconductor and dielectric materials. Consequently, thermal transport transitions from a classical diffusive regime to a mesoscopic or quasi-ballistic one, leading to localized hot spots and interfacial thermal resistance that severely restrict the operability of the device.  Understanding and accurately modeling these sub-continuum transport phenomena is therefore not merely a fundamental physics challenge, but a critical engineering need for the thermal management and architectural design of next-generation microprocessors, thermoelectric generators, and advanced quantum materials.

In the context of nanoscale thermal transport, the phonon Boltzmann transport equation (BTE) is an integro-differential equation (IDE) that constitutes the foundation of kinetic transport theory. Researchers actively leverage its asymptotic-preserving properties to design numerical frameworks that remain robust across both the ballistic-mesoscopic regime and the diffusive limit. However, directly solving the BTE presents numerous computational challenges stemming from the high dimensionality of its phase space, coupled with the intrinsic nonlinearity of the phonon-phonon scattering collision operator. To circumvent these computational problems, various deterministic and stochastic numerical schemes have been extensively developed in the literature, varying from semi-analytical methods~\cite{hua_semi-analytical_2015}, discrete ordinate methods~\cite{hu_giftbte_2023,luo_discrete_2017, zhang_fast_2021}, iterative schemes~\cite{liu_fast-converging_2022}, hybrid solvers~\cite{loy_fast_2012, sellan_cross-plane_2010}, Monte Carlo methods~\cite{pathak_mcbte_2021, mittal_monte_2010}, and, more recently, GPU-based differentiable solvers~\cite{shang_jax-bte_2025}. While effective, these traditional methods often rely on simplifying assumptions, such as linearized density functions or the relaxation time approximation based on Matthiessen's rule, sacrificing the fidelity required to accurately capture a broad range of physical phenomena. In fact, while such approximations may yield acceptable accuracy for highly symmetric crystal structures such as silicon, they fail to capture the complex, inherent anisotropy of next-generation materials. This restrictive trade-off underscores the critical need to develop highly scalable, mesh-free computational frameworks capable of resolving the multiscale transport physics without incurring prohibitive computational expenses~\cite{tran_parallel_2025}. 

Among mesh-free computational methods, physics-informed neural networks (PINNs) have emerged as powerful frameworks for the efficient solution of both forward and inverse problems governed by complex differential equations~\cite{raissi_physics-informed_2019, karniadakis_physics-informed_2021, lu_deepxde_2021}. While PINNs leverage the universal approximation properties of traditional artificial neural networks (ANNs), they differ by circumventing the reliance on massive, computationally expensive labeled datasets. Instead, PINNs are designed to act as continuous surrogate models for the underlying physical system. By embedding the governing equations directly into the network's loss function, the model can be trained iteratively to minimize the physical residual alongside appropriate boundary and initial conditions. This optimization is performed over a set of collocation points uniformly sampled across the mathematical domain of interest. By leveraging automatic differentiation (AD), spatial and temporal differential operators can be computed analytically with respect to the network's input coordinates. This capability overcomes the need for numerical approximations and grid-based discretizations, effectively circumventing the truncation errors inherent to traditional mesh-based solvers~\cite{eivazi_physics-informed_2022, chen_physics-informed_2020, chen_physics-informed_2022}.

Despite these advantages, extending standard PINN architectures to integro-differential equations such as the phonon BTE exposes their architectural limitations. Traditionally, PINN frameworks rely on deterministic numerical quadrature rules to approximate integral operators~\cite{lakshmikantham_theory_1995}. However, the discretization required for these numerical integrals naturally couples the spatial and momentum collocation points, as shown by previous attempts to solve the BTE in the literature~\cite{li_physics-informed_2021, li_physics-informed_2022, mishra_physics_2021}. This structural coupling essentially reintroduces grid-based dependencies that mesh-free methods are designed to avoid. Furthermore, it severely limits the employment of massively parallel computing architectures and stochastic mini-batch optimization, as traditional quadrature schemes demand a fixed, comprehensively evaluated set of grid points at each training iteration to guarantee the convergence of the numerical integral. 

To overcome this limitation, Yuan et al. introduced an extension of the PINN architecture known as the Auxiliary Physics-Informed Neural Network (APINN)~\cite{yuan_-pinn_2022}. Rather than relying on numerical integration, this framework expands the output space of the standard PINN to include auxiliary variables that explicitly approximate the integral term. Provided that the integral kernel satisfies specific mathematical conditions, this approach effectively recasts the governing integro-differential equation into a system of differential equations.  Consequently, the APINN formulation entirely circumvents the numerical discretization errors and computational bottlenecks associated with quadrature routines, allowing the model to evaluate all spatial and momentum operators leveraging the precision of automatic differentiation. Within the context of Boltzmann-type transport, this auxiliary method has already demonstrated remarkable efficacy in solving coupled forward and inverse problems governed by the gray model of the radiative transfer equation~\cite{riganti_auxiliary_2023}.

While the auxiliary formulation resolves the quadrature constraint associated with the integro-differential operator, it addresses only one component of the mathematical complexity of the BTE. A second, equally important computational challenge arises from the intrinsically multiscale nature of nanoscale thermal transport. In semiconductor devices, phonons exhibit temperature- and frequency-dependent mean free paths and scattering rates that span several orders of magnitude. Such a broad spectrum of carrier dynamics results in highly localized non-equilibrium behaviors and steep spatial gradients, particularly near material interfaces and localized heat sources. Traditional fully connected neural networks, however, are notoriously susceptible to spectral bias, i.e. an inherent tendency to rapidly converge on low-frequency, macroscopic features while failing to resolve high-frequency phenomena~\cite{liu_multi-scale_2020, zhang_correction_2023,riganti_multiscale_2025}. Consequently, employing a traditional single-scale PINN architecture, albeit with the auxiliary extension, will inevitably result in stiff optimization landscapes, convergence failures, and compromised physical fidelity.

To overcome the computational impediments of numerical quadrature and spectral bias, this article introduces a novel multiscale auxiliary physics-informed neural network designed to solve the generalized equation of phonon radiative transfer (GEPRT)~\cite{prasher_generalized_2003}.  Following a formal introduction of the kinetic BTE framework, the governing equations are formulated in terms of the Bose-Einstein distribution function to capture phonon dynamics. The proposed architecture is then trained to model steady-state thermal transport across a quasi two-dimensional silicon slab. To validate the proposed framework against the literature, the GEPRT is initially solved within the relaxation time approximation (RTA) using Matthiessen's rule for small temperature gradients, while the radiative equilibrium condition is explicitly enforced to train an independent, shallow neural network that resolves the effective temperature field. After this initial validation, the full GEPRT is solved for the quasi-2D silicon slab under both small and large temperature differences, and for an inverse problem. 

The computational efficiency and mesh-free nature of this framework present immediate avenues for adoption in applied engineering. In advanced semiconductor device modeling, where localized self-heating and mesoscopic thermal transport limit the performance of highly scaled transistor architectures, MTNet provides a parallelizable mechanism to evaluate and optimize thermal dissipation. Furthermore, the framework's inverse property retrieval capabilities can be directly adapted for the non-destructive characterization of thermal interface materials and the optimization of thermoelectric generators. Although demonstrated here using an isotropic scattering phase function to validate the baseline framework against established literature, the proposed architecture is inherently generalizable to fully anisotropic phase functions, as shown by previous work by the authors~\cite{riganti_auxiliary_2023}. Ultimately, MTNet establishes a robust theoretical and computational foundation for investigating kinetic transport, enabling the efficient thermal management of next-generation nanoelectronics, optoelectronics, and anisotropic materials.

\section{\label{sec:methods}Methods}
\subsection{The Boltzmann transport equation}
The Boltzmann transport equation (BTE) is an integro-differential equation that models the evolution of the seven-dimensional particle distribution function $f(\textbf{r}, \textbf{k}, t)$ across spatial, momentum, and temporal phase space.  The general governing equation is expressed as:
\begin{equation}
    \frac{\partial f}{\partial t} + \textbf{v} \cdot \nabla_{\textbf{r}} f + \textbf{a} \cdot \nabla_{\textbf{v}} f = \left[ \frac{\partial f}{\partial t} \right]_{\text{coll}}
    \label{eq:bte}
\end{equation}
where $\textbf{r}$ is the real-space position vector, $\textbf{v}$ is the particle group velocity, $\textbf{a}$ represents acceleration due to external fields, and the right-hand side denotes the collision operator, which dictates the rate of particle scattering. 

In kinetic transport physics and engineering, the BTE serves as a unifying mathematical framework for a diverse array of non-equilibrium phenomena, ranging from nanoscale thermal transport and semiconductor device modeling to atmospheric radiative transfer. However, the high dimensionality of the phase space, coupled with the often integro-differential nature of the collision operator, renders analytical solutions intractable outside of highly simplified regimes. Consequently, researchers must rely on derived approximations and sophisticated numerical schemes to accurately capture these complex transport dynamics.

A primary objective of solving the BTE is to bridge microscopic transport dynamics with macroscopic, laboratory-measurable observables. Macroscopic (and mesoscopic) properties are obtained by computing the statistical moments of the microscopic distribution function $f(\textbf{r}, \textbf{k}, t)$. For instance, in systems where energy is carried by quasi-particles characterized by a wavevector $\textbf{k}$ and a polarization or band index $p$, the macroscopic energy flux $\textbf{q}(\textbf{r},t)$ can be defined as:
\begin{equation}
    \textbf{q}(\textbf{r},t) = \sum_p \int \hbar \omega_p(\textbf{k}) \textbf{v}_p(\textbf{k}) f(\textbf{r}, \textbf{k}, t) \frac{d^3\textbf{k}}{(2\pi)^3}
    \label{eq:q}
\end{equation}
where $\hbar$ is the reduced Planck constant, $\omega_p(\textbf{k})$ is the angular frequency, and $\textbf{v}_p(\textbf{k})$ is the mode-dependent group velocity. To bridge the gap between the mathematical formulation and observables, it is often advantageous to express the transport equation in terms of the spectral intensity $I_\omega$. By grouping fundamental physical properties, including the density of states $D(\textbf{k})$, we can define the spectral intensity:
\begin{equation}
    I_\omega (\textbf{r}, \textbf{k}, t) = \frac{1}{4\pi} \hbar \omega_p(\textbf{k}) \textbf{v}_p(\textbf{k}) D(\textbf{k}) f(\textbf{r}, \textbf{k}, t)
    \label{eq:I_w}
\end{equation}
Formulating the BTE in terms of $I_\omega$ simplifies the calculation of the energy flux and other relevant measurable observables that are now obtained from the moments of the spectral intensity~\cite{van_rossum_multiple_1999}. 
\subsection{The Generalized equation of phonon radiative transfer}
In this study we consider the steady-state generalized equation of phonon radiative transfer (GEPRT) for phonon frequency $\omega$ of phonon branch $p$~\cite{prasher_generalized_2003}:
\begin{equation}
\mathbf{s} \cdot \nabla I_\omega(\mathbf{r}, \mathbf{s}) + (\kappa_\omega + \sigma_\omega) I_\omega(\mathbf{r}, \mathbf{s}) = \kappa_\omega I^0_\omega(T) + \frac{\sigma_\omega}{4\pi} \int_{4\pi} I_\omega(\mathbf{r}, \mathbf{s}') \Phi(\mathbf{s}', \mathbf{s}) d\Omega'
\label{eq:GEPRT}
\end{equation}
where $I_\omega^0 (T) = \frac{1}{4\pi} \hbar \omega_p v_{g,p} D(k) f_{BE}(\omega, T)$ is the spectral intensity associated to the Bose-Einstein equilibrium distribution function $f_{BE}$ at temperature $T$, $\Phi(\mu,\mu')$ is the scattering phase function, and $\kappa_\omega$ and $\sigma_\omega$ denote the inelastic absorption and elastic scattering coefficients, respectively~\cite{prasher_generalized_2003, howell_thermal_2020, modest_radiative_2021}.
After introducing a change of variables $x=\tilde{x}/L$ and considering a quasi-2D thin film of thickness $L$, we obtain:
\begin{equation}
    \mu \frac{\partial I_\omega}{\partial x} + \left( \frac{1}{\text{Kn}_{i,\omega}} + \frac{1}{\text{Kn}_{e,\omega}} \right)I_\omega = \frac{1}{\text{Kn}_{i,\omega}}I_{\omega}^{0}(T) + \frac{1}{\text{Kn}_{e,\omega}}  \frac{1}{2}\int_{-1}^{1}I_\omega(x,\mu')\Phi(\mu',\mu)d\mu'
    \label{eq:GEPRT1D}
\end{equation}
where $\mu$ is the cosine of the polar angle of incidence $\theta$ (assuming isotropic scattering along the azimuthal angle $\phi$), and $x$ is the dimensionless spatial coordinate. The $\omega$ subscript indicates that the GEPRT is an equation defined on a continuous frequency spectrum for a given phonon polarization branch $p$. Analogously, in the radiative transfer equation for photons, Equation~\ref{eq:GEPRT} is evaluated across an unbounded frequency spectrum~\cite{modest_radiative_2021, howell_thermal_2020}. Conversely, the phonon frequency domain for any given polarization branch is finite and upper-bounded by the maximum wavenumber in the crystal.    Consequently, solving the full phonon GEPRT requires the evaluation of a system of $p$ integro-differential equations, governed by Equation~\ref{eq:GEPRT}, that must be simultaneously satisfied at every frequency point. Eq.~\ref{eq:GEPRT1D} also introduces the frequency-dependent inelastic and elastic Knudsen numbers, which are non-dimensional parameters governing the transport regime defined as the ratio of the frequency-dependent phonon mean free path (MFP) for inelastic ($\Lambda_{i,\omega}$) and elastic ($\Lambda_{e,\omega}$) scattering events and the spatial dimension $L$:
\begin{equation}
    \text{Kn}_{i,\omega} = \frac{1}{\kappa_\omega L}=\frac{\Lambda_{i,\omega}}{L}=\frac{v_{g,\omega} \tau_{i,\omega}}{L}, \quad \quad \text{Kn}_{e,\omega} = \frac{1}{\sigma_\omega L}=\frac{\Lambda_{e,\omega}}{L}=\frac{v_{g,\omega} \tau_{e,\omega}}{L}
    \label{eq:Kn}
\end{equation}
The MFP, expressed as the product of the frequency-dependent phonon group velocity $v_{g,\omega}$ and the frequency and temperature-dependent relaxation time $\tau_\omega(T)$, represents the average spatial distance a phonon propagates through the crystal lattice before undergoing a scattering event. Inelastic scattering processes are primarily governed by anharmonic phonon-phonon interactions, encompassing both Umklapp and Normal processes, which alter the energy and momentum distribution of the phonon gas. Conversely, elastic scattering processes dictate the directional redistribution of phonons without corresponding energy dissipation. At the nanoscale, these elastic events are typically driven by isotopic impurities, point defects, and boundary scattering.

Eq.~\ref{eq:GEPRT1D} must be solved for $I_\omega$ and the effective temperature $T$. After imposing suitable boundary conditions for $I_\omega$, we close the system and retrieve the effective temperature by imposing the energy balance at every spatial point, resulting in the condition of radiative equilibrium~\cite{zhang_nanomicroscale_2020}:
\begin{align}
    &\sum_p2\pi\int_{0}^{\omega_{p}^{max}} \int_{-1}^{1} \frac{1}{\Lambda_{i}(\omega_p)} I(\omega_p,x, \mu) d\mu d\omega_p = \sum_p4\pi \int_{0}^{\omega_{p}^{max}} \frac{1}{\Lambda_{i}(\omega_p)} I^0(\omega_p,T(x)) d\omega_p \\ 
    &\sum_p 2\pi\int_{0}^{\omega_{p}^{max}} \int_{-1}^{1} \frac{I(\omega_p,x,\mu)-I^0(\omega_p,T(x))}{\Lambda_{i}(\omega_p)} d\mu d\omega_p=0
    \label{eq:rad_eq}
\end{align}
Eq.~\ref{eq:GEPRT1D}-~\ref{eq:rad_eq} with suitable boundary conditions for $I_\omega$ uniquely determine the solution of the GEPRT. For the examples that will follow, we will consider blackbody boundary conditions defined as:
\begin{align}
    I(\omega_p, x_0, \mu^+)=I^0(\omega_p,T_0) \\
    I(\omega_p, x_1, \mu^-)=I^0(\omega_p,T_1)
    \label{eq:BC_black}
\end{align}
Next, we will develop the auxiliary formulation for the GEPRT. 

\subsection{\label{subsec:aux_formulation}Auxiliary formulation of the GEPRT}
In order to recast the Boltzmann problem as a purely differential one, we rewrite Eq.~\ref{eq:GEPRT1D} and~\ref{eq:rad_eq} by introducing auxiliary differential equations for the three polarization-dependent auxiliary variables $\hat{f}_\omega,\hat{v}_\omega,\hat{g}_\omega$:
\begin{align}
    &\mu \frac{\partial \hat{I}(\omega_p,x,\mu)}{\partial x} + \left( \frac{1}{\text{Kn}_{i}(\omega_p)} + \frac{1}{\text{Kn}_{e}(\omega_p)} \right)\hat{I}(\omega_p,x,\mu) = \frac{1}{\text{Kn}_{i}(\omega_p)}I_{\omega}^{0}(\omega_p,\hat{T}(x)) + \frac{1}{\text{Kn}_{e}(\omega_p)}  \frac{1}{2}\hat{f}(\omega_p,1,x) \label{eq:aux_GEPRT1} \\
    &\sum_p 2\pi\int_{0}^{\omega_{p}^{max}} \int_{-1}^{1} \frac{\hat{I}(\omega_p,x,\mu)-I^0(\omega_p,\hat{T}(x))}{\Lambda_{i}(\omega_p)} d\mu d\omega_p=\sum_p \hat{v}(\omega_{p}^{max},\mu,x)=0  \label{eq:aux_GEPRT2} \\
    &\frac{\partial \hat{v}(\omega_p,x,\mu)}{\partial \omega_p} = \hat{g}(\omega_p,x,1) \quad \quad \frac{\partial \hat{g}(\omega_p,x,\mu)}{\partial\mu}=\frac{\hat{I}(\omega_p,x,\mu)-I^0(\omega_p,\hat{T}(x))}{\Lambda_{i}(\omega_p)} 
    \label{eq:aux_GEPRT3}\\
    &\frac{\partial \hat{f}(\omega_p,x,\mu)}{\partial\mu} = \hat{I}(\omega_p,x,\mu')\Phi(\mu',\mu)=\hat{I}(\omega_p,x,\mu), \quad\quad \Phi(\mu',\mu)=1
    \label{eq:aux_GEPRT4}
\end{align}
where:
\begin{align}
    & \hat{f}(\omega_p,\tilde{\mu},x) = \int_{-1}^{\tilde{\mu}}\hat{I}(\omega_p,x,\mu')\Phi(\mu',\mu)d\mu'\quad\quad &\hat{f}(\omega_p,x,-1)=0\\
    & \hat{v}(\tilde{\omega},x, \mu)=\int_{0}^{\tilde{\omega}} \int_{-1}^{1} \frac{\hat{I}(\omega_p,x,\mu)-I^0(\omega_p,\hat{T}(x))}{\Lambda_{i}(\omega_p)} d\mu d\omega_p\quad \quad &\hat{v}(0,x,\mu)=0\\
    & \hat{g}(\omega_p,\tilde{\mu},x) = \int_{-1}^{\tilde{\mu}} \frac{\hat{I}(\omega_p,x,\mu)-I^0(\omega_p,\hat{T}(x))}{\Lambda_{i}(\omega_p)} d\mu \quad\quad &\hat{g}(\omega_p,x,-1)=0
    \label{eq:aux_vars}
\end{align}
The hat indicates that a given function is a neural network output. For anisotropic materials, one would consider the general expansion of the scattering phase function in spherical harmonics or in Legendre polynomials~\cite{liemert_exact_2013, noauthor_modified_2003}. For instance, in dealing with the azimuthally symmetric case, we have:
\begin{equation}
    \Phi(\mu,\mu') = \sum_{\ell=0}^{\infty}w_{\ell}P_{\ell}(\mu)P_{\ell}(\mu'), \quad \quad w_\ell = (2\ell+1)g^{\ell}
\label{eq:legendre}
\end{equation}
where $g\in[0,1]$ is the factor that governs the scattering anisotropy~\cite{howell_thermal_2020, modest_radiative_2021, van_rossum_multiple_1999}.
Eq.~\ref{eq:aux_GEPRT1},~\ref{eq:aux_GEPRT2},~\ref{eq:aux_GEPRT3} convert the GEPRT to a purely differential system of equations that leverages the accuracy of automatic differentiation and GPU computing.
\begin{figure}[h]
\centering
\includegraphics[width=1\textwidth]{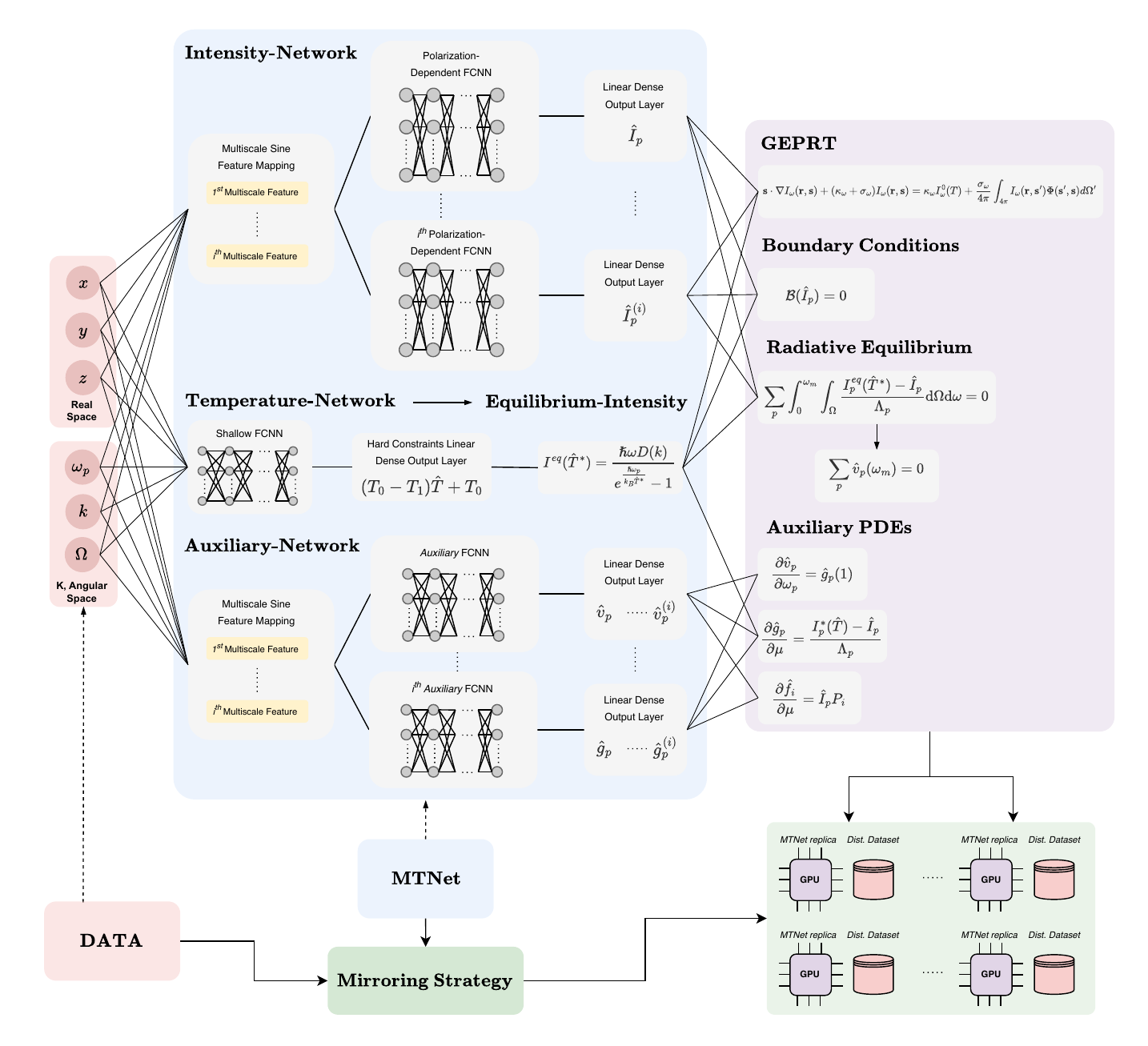}
\caption{Diagram of the MTNet architecture and schematics of the distributed training routine. The collocation points are passed to the three independent neural networks that comprise MTNet, namely the Intensity-Network, the Temperature-Network, and the Auxiliary Network. MTNet detects the available computing hardware and replicates its architecture across all available GPU devices using TensorFlow's MirroredStrategy. Then, by means of automatic differentiation (AD), the derivatives of $\hat{I}(\omega_p,x,\mu)$, $\hat{f}(\omega_p,\tilde{\mu},x)$, $\hat{v}(\tilde{\omega},x, \mu)$, $\hat{g}(\omega_p,\tilde{\mu},x)$, and $\hat{T}(x)$ along with the PDE and boundary conditions. During each training step, the global batch of collocation points is partitioned into sub-batches, and each GPU independently evaluates the network and computes the loss gradients for its designated sub-batch. The gradients are then appropriately combined and the loss functions are evaluated globally. This process is repeated until convergence.}
\label{fig:MTNet}
\end{figure}

\subsection{\label{subsec:aux_architecture}The auxiliary multiscale PINN architecture}
Figure~\ref{fig:MTNet} shows a diagram of the coupled Multiscale Thermal Network (MTNet) that was implemented to solve the auxiliary GEPRT. The spatial, angular, and k-space collocation points are passed to MTNet, which is comprised of three coupled PINNs: the Intensity-Network, the Temperature-Network, and the Auxiliary Network. The Intensity-Network is a polarization-independent multiscale PINN~\cite{riganti_multiscale_2025} with a sine feature mapping in the first layer of 64 neurons and the $x\cdot\sigma(x)$ activation function in the 3 hidden layers with 64 neurons, where $\sigma(x)$ is the sigmoid activation function~\cite{haykin_neural_2009}. The multiscale architecture is necessary to capture the sharp discontinuities in the $I_\omega$ function in the mesoscopic regime near $\mu=0$. Typically, the $I$-Network requires four to eight scales to resolve the multiscale physics in the mesoscopic regime fully. The auxiliary network implementation mirrors that of the $I$-Network, except that the auxiliary boundary conditions are embedded in the PINN output via hard constraints to reduce the number of loss function terms during training. Finally, the temperature network is a shallow 4-layer by 32 neurons PINN with hard constraints to avoid dividing by zero in the equilibrium distribution function $f_{BE}(\omega_p,T)=1/(e^{\hbar\omega_p/k_BT}-1)$ at the beginning of the training cycle. 

The outputs of the three networks are then used to compute, by means of automatic differentiation (AD), the derivatives of $\hat{I}(\omega_p,x,\mu)$, $\hat{f}(\omega_p,\tilde{\mu},x)$, $\hat{v}(\tilde{\omega},x, \mu)$, $\hat{g}(\omega_p,\tilde{\mu},x)$, and $\hat{T}(x)$ along with the PDE and boundary condition. Each calculated value is then combined into a term of the loss function $\mathcal{L}(\bm{\tilde{\theta}})$ defined as:
\begin{equation}
    \begin{split}
        \mathcal{L}(\bm{\tilde{\theta}}) = \mathcal{L}_{GEPRT}(\bm{\tilde{\theta}};\mathcal{N}_{int})+\mathcal{L}_{b}(\bm{\tilde{\theta}};\mathcal{N}_{b}) + \mathcal{L}_{aux}(\bm{\tilde{\theta}};\mathcal{N}_{aux})
    \end{split}
    \label{eq:loss}
\end{equation}
In Eq.~\ref{eq:loss}, each term has been computed using a mean-squared error (MSE) loss function, defined individually as:
\begin{align}
    \mathcal{L}_{GEPRT}(\bm{\tilde{\theta}};\mathcal{N}) =\frac{1}{|\mathcal{N}|}\sum\nolimits_{\bm{x}\in\mathcal{N}} \left|\left| f\left( \bm{x};\hat{I},\hat{f},\hat{T},\frac{\partial \hat{I}}{\partial x},\hat{v}_0,\dots, \hat{v}_n \right) \right|\right|^2_2
\end{align}
where $\mathcal{L}_{GEPRT}$ denotes the loss term calculating the GEPRT and radiative equilibrium condition,
\begin{align}
    \mathcal{L}_{b}(\bm{\tilde{\theta}};\mathcal{N}_{b}) = \frac{1}{|\mathcal{N}_{b}|}\sum\nolimits_{\bm{x}\in\mathcal{N}_{b}} \left|\left| \mathcal{B}(\hat{I},\bm{x}) \right|\right|^2_2
\end{align}
is the loss term for the boundary conditions of the GEPRT where $\bm{x}\in \partial\Omega$. Finally,
\begin{align}
    \mathcal{L}_{aux}(\bm{\tilde{\theta}};\mathcal{N}_{aux}) = \frac{1}{|\mathcal{N}_{aux}|}\sum\nolimits_{\bm{x}\in\mathcal{N}_{aux}} \left|\left| f\left( \bm{x};\frac{\partial \hat{v}_0}{\partial \omega_0},\dots, \frac{\partial \hat{v}_n}{\partial \omega_n}, \frac{\partial \hat{g}_0}{\partial \mu},\dots, \frac{\partial \hat{g}_n}{\partial \mu}, \frac{\partial \hat{f}_0}{\partial \mu},\dots, \frac{\partial \hat{f}_n}{\partial \mu} \right) \right|\right|^2_2
\end{align}
is the loss term associated with the auxiliary functions.

In the context of modern high-performance computing, formulating the GEPRT as a fully differential system unlocks substantial computational advantages, as illustrated in Figure~\ref{fig:MTNet}. Primary among these is the ability to exploit synchronous data parallelism across multi-GPU architectures. Because the auxiliary formulation decouples the collocation points dedicated to the evaluation of the integral operator, machine learning frameworks can effortlessly distribute the training workload. To do so, MTNet detects the available computing hardware and replicates its architecture across all available GPU devices using TensorFlow's MirroredStrategy~\cite{martin_abadi_tensorflow_2015}. During each training step, the global batch of collocation points is partitioned into sub-batches, and each GPU independently evaluates the network and computes the loss gradients for its designated sub-batch. These localized gradients are subsequently aggregated across the cluster via an all-reduce algorithm, ensuring that the model weights are synchronously updated. This parallelization effectively multiplies the volume of data the network processes per iteration, thereby accelerating convergence over high-dimensional phase spaces. This mesh-free approach also allows high flexibility in the allocation of computational degrees of freedom during the optimization process, enabling a fine-grained discretization of both angular and frequency space. 
For the following examples, MTNet was trained for approximately 1000 epochs on an input dataset consisting of $\mathcal{N}=\mathcal{N}_{aux}=\mathcal{N}_{b}=2^{21}\approx10^6$ total collocation points per input variable (i.e., space, frequency, and angle), which were split into $2^{11}=2048$ collocation point batches. The training routine reached a consistent six to seven orders of magnitude decrease of the cumulative loss by the end of training, capped at approximately $\mathcal{L}(\bm{\tilde{\theta}})=10^{-6}$. This synchronous data parallelism natively supports both single- and multi-GPU distributed training. To ensure broad hardware compatibility, the framework was validated across diverse NVIDIA architectures, specifically the V100 (compute capability 7.0), L40S (compute capability 8.9), and H200 (compute capability 9.0). Consequently, MTNet is not strictly reliant on massive high-performance computing (HPC) clusters, offering a highly scalable solution that remains accessible to standard laboratory workstations.

\section{\label{sec:results}Results}
We consider the cross-plane phonon transport in a silicon thin film of thickness $L$ heated by two blackbodies located at its left and right boundaries and maintained at constant temperatures $T_0$ and $T_1$, respectively. To facilitate the comparison with previously published results on phonon BTE~\cite{luo_discrete_2017}, monocrystalline silicon with a quadratic dispersion of the longitudinal acoustic (LA) and doubly degenerate transverse acoustic (TA) phonon branches was considered:
\begin{equation}
    \omega_p = c_{p0}k + c_{p1}k^2, \quad \quad v_{g,p}=\frac{\partial \omega_p}{\partial k}
\end{equation}
Where $k=0.5431$ nm for silicon, assuming an isotropic crystal, and $D(k)=\frac{k^2}{2\pi^2 v_g}$. The parametrized scattering rates and silicon band parameters are taken from Ref.~\cite{luo_discrete_2017,hopkins_reduction_2011}. 
\begin{figure}
\centering
\includegraphics[width=1\textwidth]{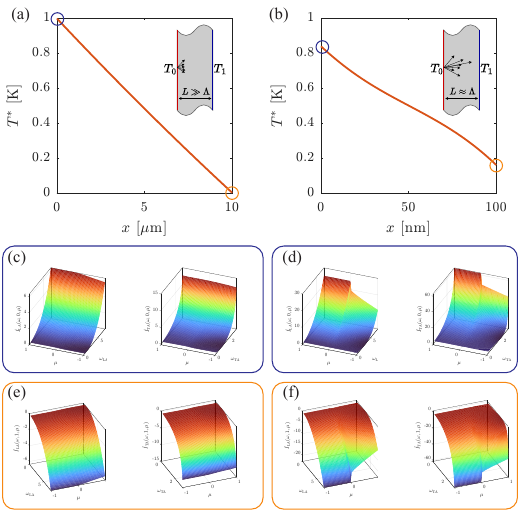}
\caption{(a-b) Solution of the EPRT in the relaxation-time approximation for the diffusive and mesoscopic regimes, respectively, in the silicon thin film geometry schematically shown over the plots. The black arrows are meant to schematically represent a few phonon scattering lengths $\Lambda$, which are much smaller than $L$ for diffusive transport, and of the order of $L$ for mesoscopic transport. The blue and orange circles correspond to the respective boxes in panels (c), (f), where the phonon intensity has been plotted at a single spatial point as a function of frequency and angle. It can be noticed that, in the intensity plots for the mesoscopic regime shown in panels (d) and (f), the intensity displays a sharp asymmetry about $\mu=0$. This asymmetry is almost absent in the intensity plots in panels (c) and (e), which is expected in the diffusive regime.}
\label{fig:RTA_smallDT}
\end{figure}
\subsection{\label{subsec:RTA}Validation of the GEPRT in the relaxation time approximation}
To validate MTNet against previously published results, we first consider the relaxation time approximation (RTA) for a small temperature difference of $T_0-T_1=\Delta T=1\rm K$:
\begin{equation}
    \mu \frac{\partial I(\omega_p,x,\mu)}{\partial x} = \frac{1}{\text{Kn}(\omega_p)}(I^{0}(\omega_p,T)-I(\omega_p,x,\mu))
    \label{eq:RTA}
\end{equation}
In Eq.~\ref{eq:RTA}, known as the equation of phonon radiative transfer (EPRT)~\cite{zhang_nanomicroscale_2020}, the individual relaxation times were combined in a single relaxation time using Matthiessen's rule, resulting in a single frequency-dependent Knudsen number $\text{Kn}(\omega_p)$~\cite{zhang_nanomicroscale_2020, prasher_generalized_2003, joshi_transient_1993}. For this first example, the auxiliary system of equations~\ref{eq:aux_vars} is simplified to include only $\hat{v}$ and $\hat{g}$ because the RTA does not include the scattering phase function from~\ref{eq:aux_GEPRT1}. However, no further approximation is employed, and the Temperature network is trained at each epoch by satisfying the condition of radiative equilibrium, i.e., Eq.~\ref{eq:rad_eq}, explicitly. 
Results for this first simulation are shown in Figure~\ref{fig:RTA_smallDT}, where panels (a) and (b) were obtained for two different film thicknesses $L$, matching the well-known literature results~\cite{hua_semi-analytical_2015, luo_discrete_2017, liu_fast-converging_2022}. Panel (a), obtained by choosing $L=10\mu \rm m$, displays the linear temperature profile that is characteristic of the diffusive regime, i.e., when the average phonon MFP $\Lambda \ll L$ and the average of the Knudsen number is much smaller than unity. On the other hand, panel b displays the signature temperature slips in the temperature solution that are characteristic of the mesoscopic transport regime when $\Lambda\approx L=100 \rm{nm}$ and $\rm{Kn}\approx1$. In Figure~\ref{fig:RTA_smallDT}, we have also included the boundary contour plots for $\hat{I}$ in panels (c)-(f), which are color-coded to match the circles in (a)-(b). As expected, it can be noticed that the intensity contour plots in the diffusive regime, namely (c) and (e), are almost symmetric about $\mu=0$ at both boundaries. This symmetry is expected because, in the diffusive regime, scattering events in the bulk occur at a high enough rate that the out-of-equilibrium phonon intensity relaxes very close to the value of equilibrium intensity $I^0$ injected into the system by the blackbodies at the boundary. The slight asymmetry, noticeable for negative $\mu$ values, is necessary for the constant heat flux to be present in a system without an internal heat source. On the other hand, the boundary plots for the mesoscopic regime displayed in panels (d) and (f) display a sharp asymmetry about $\mu=0$. This feature is also expected because the relatively similar size of the device dimension to the phonon MFP splits the transport regime between a weaker scattering bulk transport regime, characterized by a flatter temperature curve, and a diffusive boundary transport characterized by the linear regime from panel (a). The combination of the two transport regimes results in a steady state temperature profile that "slips" at the boundaries, which validates previously published results in this regime~\cite{hua_semi-analytical_2015, luo_discrete_2017,liu_fast-converging_2022}. To confirm the validity of the solution further, we compute the average of the effective thermal conductivity $k_{\text{eff}}=\textbf{q}L/\Delta T$ in both examples, retrieving the literature values of $k_{\text{eff}}/k_{\text{bulk}}\approx1$ and $k_{\text{eff}}\approx30$ W/m/K, $k_{\text{eff}}/k_{\text{bulk}}\approx0.2$ in the diffusive and mesoscopic regime, respectively~\cite{luo_discrete_2017}. 
\begin{figure}[h]
\centering
\includegraphics[width=1\textwidth]{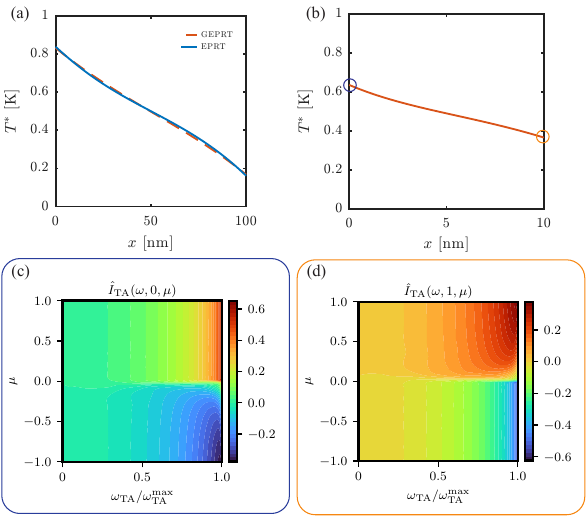}
\caption{(a) Agreement between the temperature profiles of the GEPRT and RTA-based EPRT at small temperature differences. (b) GEPRT prediction of the temperature profile in the weakly scattering regime for silicon. The blue and orange dots correspond to the colored boxes in panels (c) and (d), where a representative phonon intensity for the TA polarization has been plotted as a function of frequency and angle. The two panels show the precision of the multiscale architecture in capturing the discontinuity at $\mu=0$.}
\label{fig:comp}
\end{figure}
\subsection{GEPRT solutions for cross-plane phonon transport}
Next, MTNet is trained to solve the full GEPRT from Section~\ref{subsec:aux_formulation} for a silicon thin film. To compare the results with the RTA from Section~\ref{subsec:RTA}, we utilize the same phonon parameters and set the scattering phase function to $\Phi(\mu',\mu)=1$, which is a valid scattering model for monocrystalline silicon. This model can be naturally extended to anisotropic materials by including the suitable anisotropy term $g$ governing preferred scattering directions~\cite{van_rossum_multiple_1999, riganti_auxiliary_2023} and a directional group velocity $\textbf{v}_g$. 

The fundamental physical distinction between the EPRT under the relaxation time approximation (RTA) and the full GEPRT lies in their respective treatments of the elastic scattering coefficient, $\sigma_\omega$, governed by the elastic Knudsen number $\text{Kn}_{e}$. In the standard RTA framework, the entire collision operator relaxes the non-equilibrium distribution uniformly towards the local equilibrium intensity, $I^0(T)$. Consequently, this approach mathematically combines elastic impurity scattering with inelastic Umklapp and Normal processes, erroneously treating all scattering events as fully dissipative mechanisms~\cite{chen_nanoscale_2023, chiloyan_greens_2021}. Conversely, the GEPRT explicitly enforces energy conservation during elastic processes, and phonons undergo purely directional redistribution while retaining their original spectral energy. Because these elastically scattered heat carriers continue to contribute to the net thermal flux across the domain rather than being artificially thermalized, the GEPRT inherently predicts a larger, more physically accurate effective thermal conductivity compared to the RTA-based EPRT.
Figure~\ref{fig:comp}(a) compares the computed cross-plane temperature profiles for both models. Although the two solutions exhibit near-perfect agreement in their spatial temperature distributions, the GEPRT yields a notably higher effective thermal conductivity of $k_{\mathrm{eff}} = 42~\mathrm{W/m/K}$ compared to the standard EPRT. The equivalence of the temperature profiles in Figure~\ref{fig:comp}(a) occurs because the profile is primarily affected by high-frequency phonons that thermalize rapidly in both frameworks, effectively locking the temperature gradient in place~\cite{dames_theoretical_2004}. The divergence in conductivity arises because the GEPRT properly conserves the energy of fast, low-frequency phonons during elastic isotope scattering, allowing them to transport a significantly larger total heat flux than the RTA model, which artificially suppresses flux by forcing all modes toward a local equilibrium bath~\cite{chiloyan_greens_2021}. 

Figure~\ref{fig:comp}(b) shows an additional temperature profile obtained in a weakly scattering regime at $L=10\rm{nm}$. In this solution, the temperature slips appear even more pronounced, approaching the fully ballistic or acoustically thin limit~\cite{zhang_nanomicroscale_2020}. To demonstrate the precision of the multiscale architecture, Figures~\ref{fig:comp}(c) and (d) present representative phonon intensity profiles for the transverse acoustic mode at both spatial boundaries. As the system approaches the ballistic limit, the angular discontinuity at $\mu=0$ becomes increasingly pronounced, transitioning from the slight asymmetry observed in Figures~\ref{fig:RTA_smallDT}(c) and (e) to the step-function-like profile displayed in Figure~\ref{fig:comp}. This weakly scattering regime is precisely where the multiscale formulation excels~\cite{riganti_multiscale_2025}. By effectively resolving the fast-varying Fourier components of the distribution near the singular point, the architecture yields a highly accurate solution that predicts a severely suppressed effective thermal conductivity of $k_{\text{eff}} \approx$ 7 W/m/K.
\subsection{GEPRT solutions under large temperature gradients}
Building upon this initial validation, the MTNet framework is now extended to simulate cross-plane phonon transport across a silicon thin film of thickness $L$ subjected to a large temperature gradient. Historically, this highly non-equilibrium regime has been sparsely addressed in the literature within the context of the RTA-based EPRT. This omission stems primarily from the mathematical breakdown of the standard linearization method applied to the equilibrium Bose-Einstein intensity, an approximation that breaks down at large $\Delta T$~\cite{hua_semi-analytical_2015, tran_parallel_2025, luo_discrete_2017,liu_fast-converging_2022}. In addition, the physical parametrization of the inelastic scattering coefficients exhibits a strong, explicit temperature dependence. Consequently, these coupled effects render the governing integro-differential formulation in Equation~\ref{eq:GEPRT1D} highly nonlinear, posing a challenge for traditional numerical solvers. Previous efforts to resolve large temperature gradients within the phonon BTE using physics-informed neural networks were reported by Li et al.~\cite{li_physics-informed_2022}. However, their room-temperature demonstrations involving large $\Delta T$ were strictly confined to the macroscale diffusive regime, where transport is adequately governed by the classical diffusion equation. Their framework pretrains a shallow neural network to function as an offline algebraic interpolator. This sub-network is trained to minimize the residual against the analytical Bose-Einstein distribution, generating a temperature-dependent scaling factor, $\beta(T)$, that recasts the equilibrium distribution as a pseudo-linear expansion: $f^{\text{eq}}(T) \approx f^{\text{eq}}(T_{\text{ref}}) + \beta(T)(T - T_{\text{ref}})$. Crucially, because this pretraining routine relies exclusively on isolated macroscopic temperature differentials and entirely omits spatial dependencies, it fundamentally decouples the scattering collision operator from the local spatial gradients. While this shallow network effectively circumvents the immediate quadrature limitations, it artificially constrains the scattering physics. Consequently, the architecture structurally biases the solution toward smooth, diffusive profiles and thereby suppresses the sharp, spatially coupled kinetic phenomena necessary to capture mesoscopic transport. In contrast, the auxiliary formulation of MTNet explicitly preserves and evaluates the exact, fully nonlinear integro-differential operators continuously across the spatial domain without resorting to surrogate linearizations. In this section, after an initial validation of the proposed architecture within the macroscopic diffusive limit, MTNet is trained to resolve the mesoscopic temperature field under a large temperature difference at room temperature.
\begin{figure}
\centering
\includegraphics[width=1\textwidth]{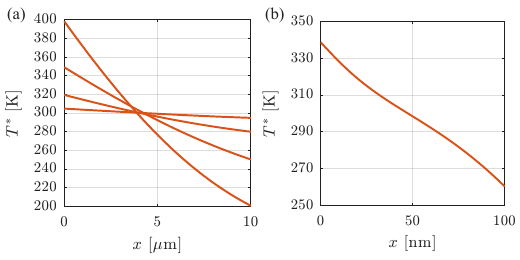}
\caption{(a) Large $\Delta T$ temperature solutions for the GEPRT in the diffusive regime. As expected due to explicit temperature dependence in the scattering rates, the large thermal gradient introduces a noticeable asymmetry in the temperature solution. (b) Temperature solution at large $\Delta T$ in the mesoscopic regime. The solution exhibits the characteristic boundary temperature slips and S-shaped" curvature indicative of the mesoscopic transport regime.}
\label{fig:RTA_largeDT}
\end{figure}
Figure~\ref{fig:RTA_largeDT}a shows the temperature profiles obtained by training MTNet at varying $\Delta T=10\rm K,20\rm K,100\rm K, 200\rm K$ for a silicon slab of thickness $L=10\mu\rm m \gg \Lambda$. As expected in systems with large temperature gradients, this transport exhibits a visible spatial asymmetry. Because inelastic scattering rates vary with the local temperature, the phonon mean free path continuously changes across the domain. Consequently, transport near the hot boundary is highly collisional and diffusive, resulting in a steep temperature gradient, while the cold boundary experiences elongated mean free paths, driving the system into a more ballistic regime characterized by a flatter temperature profile. In the macroscopic limit where boundary slips vanish, this asymmetric curvature is often captured by a heat equation employing a spatially-dependent thermal conductivity $k(x)$, which yields a similar non-linear power-law temperature decay~\cite{li_physics-informed_2022}.

Next, we train MTNet to simulate mesoscopic phonon transport across a 100 nm silicon thin film subjected to a large temperature difference of $\Delta T = 100$ K. The predicted spatial temperature distribution, presented in Figure~\ref{fig:RTA_largeDT}b, distinctly captures the characteristic thermal boundary slips while maintaining the symmetric, non-linear "S-shaped" curvature indicative of the mesoscopic transport regime. Because the extreme non-equilibrium nature and strong nonlinearities of this large-$\Delta T$ regime currently preclude direct validation against traditional numerical BTE solvers, we evaluate the physical consistency of our model through its optimization trajectory. During training, the composite physical loss function of MTNet monotonically converged, decreasing by approximately four orders of magnitude. This strict minimization of the governing integro-differential residuals provides strong quantitative confidence in the mathematical convergence and physical fidelity of the predicted solution, demonstrating MTNet's capability to operate robustly beyond the limits of standard numerical approximations.
\begin{figure}[h]
\centering
\includegraphics[width=1\textwidth]{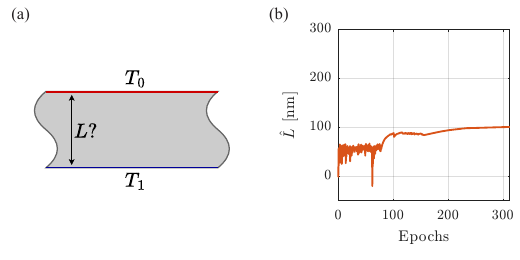}
\caption{(a) Diagram of the inverse problem, consisting of a fabricated quasi-2D silicon thin film. To infer the effective thermal transport length $L$, the inverse solver utilizes incident blackbody intensities as the driving potentials, constrained by the instantaneous lattice temperatures at the boundaries. This setup directly simulates a non-destructive metrology environment where a laboratory instrument can only probe the surface temperature of the silicon lattice. (b) Asymptotic convergence of MTNet to the effective thermal transport length $L$ that corresponds to the measured temperature slip.}
\label{fig:inv}
\end{figure}
\subsection{Inverse problems with the GEPRT}
As a final demonstration of the capabilities of the proposed framework, we deploy MTNet to solve a geometric inverse problem. Despite demonstrating high fidelity in forward simulations, physics-informed neural networks should not currently be viewed as outright replacements for traditional numerical solvers. Rather, their capacity to rapidly evaluate complex parametric spaces without grid-based constraints positions them as highly effective preconditioners, providing computationally inexpensive, physics-grounded initializations that can significantly accelerate the convergence of high-precision deterministic solvers. In both industrial and applied research contexts, deploying these predictive surrogate models can drastically reduce the reliance on computationally exhaustive, brute-force prototyping. Notably, while recent advances in data-driven scientific machine learning, such as Fourier neural operators (FNOs)~\cite{lu_learning_2021, li_fourier_2021} and diffusion-based models~\cite{bastek_physics-informed_2025}, offer unprecedented inference speeds, these architectures fundamentally lack the rigorous physical constraints required to generalize accurately in regimes where training data is scarce or nonexistent. Physics-informed neural networks circumvent this limitation, and by functioning as unsupervised continuous surrogates, their architectures inherently parameterize the solution space using adaptive, nonlinear basis functions. This adherence to the governing physical equations renders PINNs an ideal bridge, closing the gap between the rigorous fidelity of traditional numerical solvers and the computational agility of modern scientific machine learning~\cite{lu_deepxde_2021,chen_physics-informed_2020, chen_physics-informed_2022, lu_physics-informed_2021,yuan_-pinn_2022, riganti_auxiliary_2023,riganti_multiscale_2025, riganti_ddnet_2026}. In the context of nanoscale thermal transport and its associated inverse problems, the development of continuous, mesh-free solvers is imperative to overcome the limitations of traditional approximations and recover high-fidelity, physics-based sub-continuum phenomena. Within the broader scope of nanoscale inverse problems, the accurate retrieval of the effective phonon scattering length is critical for the thermal characterization of semiconductor structures. This parameter rigorously determines the mesoscopic thermal resistance and the ballistic transport limits in thin films. However, standard experimental metrology, such as time-domain thermoreflectance (TDTR) and Raman thermometry, is limited to surface- or interface-level thermal measurements~\cite{saurav_anisotropic_2024,tran_parallel_2025}. Consequently, the ability to extract internal structural parameters from localized boundary data is essential to bridge computational models with physical metrology. To demonstrate the practical utility of the proposed framework in this context, we formulate a specific geometric inverse problem aimed at recovering the unknown effective thermal transport length utilizing only steady-state interface temperatures.

The physical configuration, illustrated schematically in Figure~\ref{fig:inv}(a), consists of a quasi-2D silicon thin film. To extract the thermal transport length $\hat{L}$ corresponding to the imposed boundary temperatures, the physical boundary conditions of the transport model are specified by incident blackbody intensities. Simultaneously, the framework is provided with the local phonon gas temperatures at the boundaries, simulating the observational data of a thermoreflectance probe capable of measuring only the instantaneous lattice temperature at the film surfaces. The discrepancy between the driving emissive walls and the measured lattice temperatures provides the necessary slip dynamics to inversely solve for $L$.
% The physical configuration, illustrated schematically in Figure~\ref{fig:inv}(a), consists of a fabricated quasi-2D silicon thin film. To understand the effective thermal transport length $L$, the boundary conditions are fixed at $T_0$ and $T_1$, simulating the boundary readings of a laboratory instrument that can only probe the instantaneous temperature of the silicon lattice at the boundaries.
% The physical configuration, illustrated schematically in Figure~\ref{fig:inv}(a), consists of a four-layer heterostructure material stack. The objective is to retrieve the unknown geometrical thickness, $L$, of the intermediate layer ($Mat_2$) required to sustain specific steady-state thermal boundary conditions: a fixed temperature $T_0$ at the $Mat_1/Mat_2$ interface, and $T_1$ at the $Mat_2/Mat_3$ interface. 
In practice, this is achieved by adding a loss function term in Eq.~\ref{eq:loss}, which is modified to include the two desired temperature data points at the boundaries of the thin film:
\begin{equation}
    \begin{split}
        \mathcal{L}(\bm{\tilde{\theta}}) = \mathcal{L}_{GEPRT}(\bm{\tilde{\theta}};\mathcal{N}_{int})+\mathcal{L}_{b}(\bm{\tilde{\theta}};\mathcal{N}_{b}) + \mathcal{L}_{aux}(\bm{\tilde{\theta}};\mathcal{N}_{aux}) + \mathcal{L}_{inv}(\bm{\tilde{\theta}};\mathcal{N}_{inv})
    \end{split}
    \label{eq:loss_inv}
\end{equation}
where:
\begin{equation}
    \mathcal{L}_{aux}(\bm{\tilde{\theta}};\mathcal{N}_{aux})= 
    \frac{1}{2}\left(\left|\left| \hat{T}(0) - T_0\right|\right|^2 + \left|\left| \hat{T}(1) - T_1\right|\right|^2\right)
\end{equation}
Finally, the GEPRT needs to be modified to include the free parameter $\hat{L}$, which is now added to the output space of MTNet:
\begin{equation}
    \mu \frac{\partial \hat{I}(\omega_p,x,\mu)}{\partial x} + \left( \frac{\hat{L}}{\Lambda_{i}(\omega_p)} + \frac{\hat{L}}{\Lambda_{e}(\omega_p)} \right)\hat{I}(\omega_p,x,\mu) = \frac{\hat{L}}{\Lambda_{i}(\omega_p)}I_{\omega}^{0}(\omega_p,T) + \frac{\hat{L}}{\Lambda_{e}(\omega_p)}  \frac{1}{2}\hat{f}(\omega_p,1,x) \label{eq:aux_GEPRT_inv}
\end{equation}
While a conceptually similar inverse property retrieval was recently investigated by Shang et al.~\cite{shang_jax-bte_2025}, their differentiable BTE solver required the provision of sparse temperature data sampled from the interior of the domain. In practical laboratory settings, acquiring high-fidelity spatial temperature profiles within the bulk of sub-micron thin films is notoriously difficult. By contrast, our architecture successfully recovers the unknown thickness parameter in the mesoscopic transport regime, shown in Figure~\ref{fig:inv}(b), relying exclusively on interface-level temperature constraints. This example employs the temperature values $T_0$ and $T_1$ taken from the forward solution in Ref.~\cite{luo_discrete_2017} in order to test MTNet's capabilities against a known result. The inverse problem did not add computational overhead to the calculations, compared to the forward problem.
The practical implications of this capability are highly consequential for nanoscale materials characterization, as the developed framework can be naturally extended to the case of non-isotropic phase functions. Moreover, standard experimental metrology techniques, such as time-domain thermoreflectance (TDTR)~\cite{saurav_anisotropic_2024}, are inherently restricted to probing accessible surfaces or interfaces. By eliminating the dependence on volumetric temperature data, MTNet can be directly coupled with these non-destructive measurements. Consequently, this interface-only constraint renders this differentiable GEPRT framework a deployable diagnostic tool. This facilitates the direct extraction of thermal properties and hidden geometrical parameters, offering a mechanism for evaluating and optimizing thermal management in complex semiconductor materials. In the context of multilayer heterostructures, future MTNet extensions will incorporate interface models such as thermal boundary resistance (TBR) to accurately capture interface phonon transmission and reflection across different materials~\cite{zeng_phonon_2000, chen_thermal_1998, minnich_quasiballistic_2011}.

\section{Conclusions}
In this work, we introduced a multiscale auxiliary physics-informed neural network (MTNet) to solve the generalized equation of phonon radiative transfer. By employing an auxiliary formulation, we recast the GEPRT into a purely differential system. This framework circumvents the computational restrictions associated with quadrature rules, enabling the analytical evaluation of the integro-differential collision operator via automatic differentiation and facilitating scalable, multi-GPU parallelization. We demonstrated the physical fidelity of this architecture by simulating cross-plane mesoscopic thermal transport in a silicon thin film. Unlike traditional approximations that fail under non-equilibrium conditions, MTNet successfully captured the mesoscopic transport regime and characteristic thermal boundary slips across a large temperature gradient ($\Delta T = 100$ K). Furthermore, by explicitly distinguishing elastic from inelastic scattering mechanisms, this GEPRT implementation preserves the net thermal flux contributions of directionally redistributed phonons, overcoming the artificial thermalization inherent to the standard relaxation time approximation based on Matthiessen's rule. Finally, we demonstrated the ability of MTNet to accurately solve the inverse parameter retrieval problem of geometric inversion. The network successfully recovered the mesoscopic thermal transport length $\hat{L}$ relying exclusively on interface-level temperature constraints. By eliminating the requirement for interior bulk measurements, this approach aligns the capabilities of differentiable BTE solvers with realistic experimental constraints, paving the way for applications to heterostructure engineering and non-destructive thermal metrology. The present study establishes the power and accuracy of the MTNet architecture, introducing a highly scalable, fully differentiable paradigm for nanoscale thermal modeling, which is critical for the design and optimization of advanced nanomaterials, including nanocomposites, nanofluids, nanofiber-laden systems, and thin-film thermoelectrics~\cite{prasher_phonon_2003}. Beyond thermal transport, the GEPRT framework can naturally be extended to the electron transport by coupling the Boltzmann transport equation for the electron distribution function (and electron temperature $T_e$) with a suitable phase function for advanced semiconductor device simulations~\cite{zhang_nanomicroscale_2020,prasher_generalized_2003, li_physics-informed_2023}. This adaptability establishes a rigorous pathway to resolve complex nanoscale electron-phonon coupling mechanisms without relying on classical approximations. Accurately capturing these highly non-equilibrium coupled dynamics is a fundamental prerequisite for the design and characterization of next-generation nanoelectronic devices.
% To address these complex applications, ongoing efforts are extending the MTNet framework to accommodate generalized spatial dimensions ranging from macroscopic wafers to quasi-1D nanotubes, as well as anisotropic phase functions and hybrid Fourier-BTE multiscale simulations.

\section{Acknowledgments}
L. D. N. acknowledges the support from the U.S. Army Research Office, RF-Center managed by Dr. T. Oder (Grant \#W911NF-22-2-0158). The authors thank Professor Enrico Bellotti, Professor Alan McGaughey, and Professor Wei Cai for insightful discussions. Numerical simulations were performed on the Shared Computing Cluster (SCC) at Boston University.

%Bibliography
\bibliographystyle{unsrt}  
\bibliography{MTNet}  

\end{document}